# Convert Monolithic Application to Microservice Application

Hatem Hammad[1], Thaer Sahmoud[1], Abed Al Rahman Abu Ghazala[1]

[1]Computer Engineering Department, Islamic University of Gaza, Palestine

**Abstract**- Microservice architecture is a trending topic in software design architecture and many enterprises adopted microservice design due its benefits and the rapid and wide deployment of cloud computing and as a result, many enterprises transformed their existing monolithic application to microservice to achieve business requirements as scaling up and agile development. In this paper we will guide software developers how to convert their existing monolithic application into microservice application without re-writing the whole microservice application from scratch, and we will also discuss the common issues that may face the software developer during the conversion processes. In addition to converting the business logic to microservice, we mention steps for converting the monolithic database into a database per service. Also, we summarize how Netflix and Airbnb converted their monolithic application to microservice application.

**Keywords-** *Monolithic, Microservice, Serverless, Refactor database, Netflix.*

## 1. INTRODUCTION

There are mainly two software design approaches, the traditional software design approach which is monolithic model [1] and the modern design approach which are microservices model [2] and serverless model [3], the word monolithic is an ancient Greek word mean the large single stone, the monolithic architecture design is considered as the traditional way to build an application where the application is designed as a single codebase with a single database. The monolithic application is usually divided into three user interface layer, business logic layer and data access layer as shown in figure 1. The user interface layer is the front-end layer which interacts with users by displaying required information to user and interpret user input commands, the business logic layer is responsible for execute business functions and it process data received from the user interface layer and also interacts with data access layer, the third layer is data access layer which store data and provide access to the stored data.

Since the monolithic application is built as a single individual application then it's easy to be deployed and monitored, and also, we can simply test the end-to-end system performance.

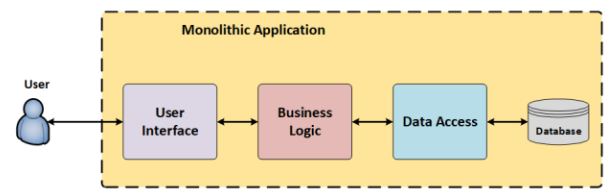

*Figure 1: Monolithic Application*

On the other hand, it's hard to develop the application when the code become so big, and in case of making any change on the application that means to redeploy the whole application and if we need to scale up any code module, we can only scale up the entire system, and it's complex to debug the code in case of error detection which is the issue that faced Netflix on their monolithic application when the whole system went down for three days because of the missing of only one semicolon on the application's code, do on other words on monolithic application the whole system would be up and running or the whole will be down and crashed. Also, it's very difficult to adopt new technologies on monolithic applications because the whole application is written as a single codebase.

Because of these drawbacks of monolithic architecture and the rapid development in cloud computing [4], software architectures split the monolithic application into multi-service architecture named as microservice. Microservices architecture in which the application is built of loosely coupled services each running on its own process and communicating with each other's API [Masse, Mark]. microservice application is usually built using API gateway approach [5] as shown in figure 2, the API gateway main task is to handle user requests and route it to the desired service, the API gateway may invoke multiple services for a single user request and aggregate the results from microservice then translate it to the user using HTTP protocol [6] as example. Each service in microservice design has its own database which can accessed using its microservice only and any other service can access only other databases via it's API.

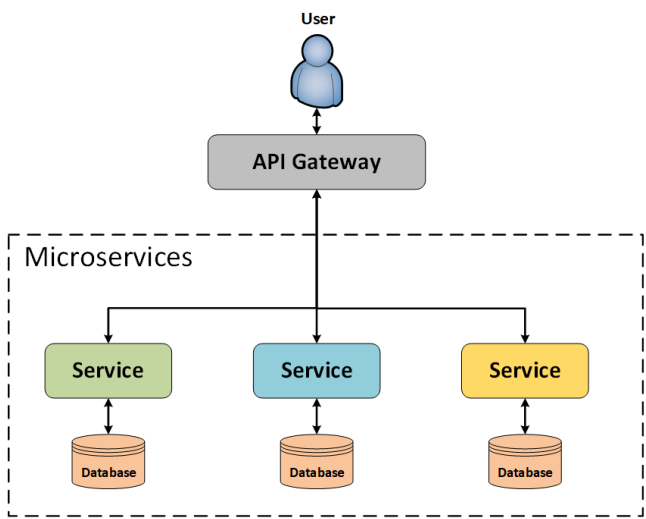

*Figure 2: Microservice Architecture*

The main benefits of using microservice architecture design is that by dividing the application into multi-services it allows developers to simply develop and understand the services, and since each service is deployed on a single host then it's easy to deploy new technologies. Also, microservice design enables us to scaling service independently, besides that we can apply changes to a service without affecting the entire application so that enables us to easily rollback any service with minimal cost.

These benefits of using microservice architecture do not come for free, cause using microservices increase the complexity of the application since it's distributed over many microservices and the communication between the services is a complex task that have to be done carefully, the design also requires skilled engineers to develop, deploy and manage the application, in addition, in addition it's hard to test end-to-end application performance. Adopting microservice is more expensive than the monolithic application since the more the services the more the resources required and hence the more the cost.

Serverless model -also known as Function as a Service FaaS- is a cloud-computing model where the application is split into functions instead of services and these functions are hosted on a cloud provider and these functions are executed when a trigger event like HTTP request, database update or file download is happened and once the function is executed it turns off until the new trigger. Serverless does not mean that servers do not exists, but servers' management and control are cloud provider responsibility and not developer responsibility, so software developers are focusing on the application itself only, and as a result serverless model is easy to be deployed and it's also a cost efficient since enterprises pay only when functions are executed, all these benefits make serverless application a suitable choice for start-up applications. Amazon Lambda [7], Google Cloud functions [8] and Microsoft Azure functions [9] are examples for FaaS providers. Microservice can be written as event-driven functions and that's a good choice when the service needs to be run occasionally, this model is often called "Serverless Microservice".

The rest of this paper is constructed as follow, in the next section we illustrate from where to start the migration and how to achieve microservices advantages on the existing monolithic application without moving to microservices approach, on section three we show methods to tear down the monolithic code base into small services, and on section four we cover how to decompose the database into a database for each service. We take a Netflix and Airbnb migration from monolithic to microservice as a case study on section five, and finally section six is the conclusion where we discuss some common issues that developers have to avoid during the migration.

## 2. WHERE TO START THE MIGRATION

As discussed in the previous section that when the application growth it leads to more complexity to understand the code and more difficulty to debug errors in addition to all the mentioned drawbacks of monolithic architecture, because of these drawbacks many enterprises like Netflix, Amazon, ebay and Spotfy convert their traditional monolithic application to microservice architecture.

So the first thing that need to be clear for making the decision of migration from monolithic to microservice is what is the goal of the migration and what's the feature you hope to achieve, because some of microservices pros can be achieved partially on monolithic application without moving to microservice architecture, as example if the target is to make a scalable system we can make vertical scaling of the monolithic application by more resources for the monolithic on the same machine or we can also do a horizontal scaling up [10] by deploy multiple copies of the monolithic and use a load balancer to distribute traffic between the copies, that approach could increase the reliability of the system since if one monolithic went down the other copies will keep working, also we can adopt new technologies on monolithic design if it's rum on Java Virtual Machine since we can run code written in different languages with in the same running process. these approaches can be thought as a short-term solution to give monolithic some of microservices benefits partially, however if the goal of the migration is clear for the software developers, then they need to incrementally accomplish the migration using small steps so that you can easily rollback in case of anything went wrong during the migration and as a result it will reduce the cost of the error. In the next section we will illustrate some approaches to convert the code base into microservice.

## 3. DECOMPOSITION OF THE CODE BASE

To transform an application from the legacy monolithic application to microservice, you do not need to re-write the application from scratch using microservice architecture because in most cases that will not work. Instead, we can convert the existing monolithic to microservice application, and to do so, firstly we need to avoid adding any new functionality to the monolithic application, secondly we need to split the user interface "front-end" from the business logic "back-end" and finally we need to decomposed and the monolithic functionality in incremental method using the Domain Driven Design "DDD" [10], we start the "DDD" and by analysing the business domain to understand the application's functional requirement, then we define the bounded context of the domain where each bounded context represent a domain model that's a subdomain of the monolithic application, in figure 3 we have an example of "DDD" bounded context for a monolithic application, each bounded context is a potential service to be extracted out from the monolithic application, and since it's recommended to start with small service, then we need to select the service with less incoming and outcoming independencies so if there's anything went wrong in extraction or even after the extraction it will have less effect on the application, in addition it requires a minimal work to be extracted, so based on that it's clear that service "A" is a good choice to be extracted from the monolithic, while extraction of service "E" requires too many work to be done because of all the incoming and outcoming decencies, so calls to that service have to be changed from local calls to service calls using APIs.

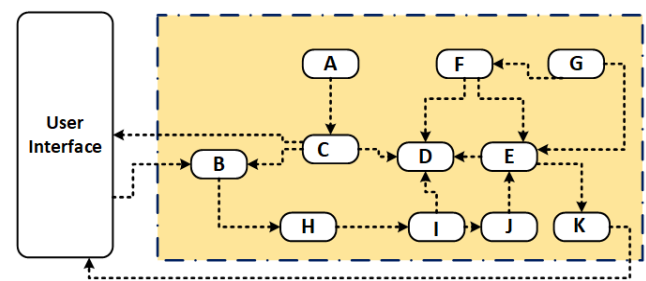

*Figure 3: Domain Driven Design of a Monolithic application*

After extracting the required service from the monolithic and changing the call of that service from being a local call to service call via API we then can apply a strangler fig pattern [12] as in figure 4. From strangler application we can see that the extracted service "A" is connected to the monolithic application via a glue code which is responsible for data integration between the extracted service and the monolithic application and the service use that glue code to read and write data owned by the monolithic, the glue code sometimes named anti-corruption layer because it ensures the new service is not polluted by data models required by the monolithic application. The strangler fig application also adds a new layer between the user interface and the business logic which is the API gateway which has the main rule of routing requests to the desired part of application (to the new extracted service or to the legacy application)

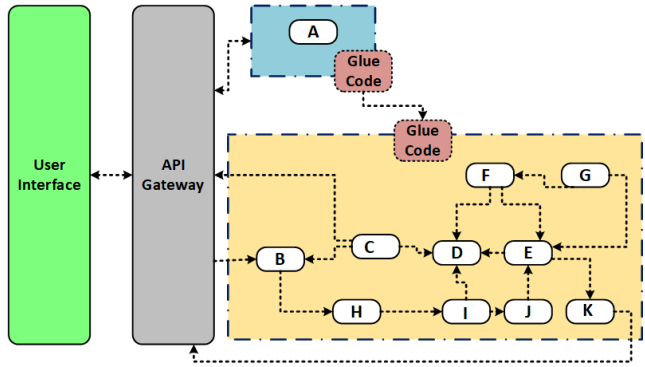

Figure 4: Strangler Fig pattern

Up to this point, both the new service and the legacy monolith application have one common database as shown in figure 5, and the extracted service can access data from the monolithic application using some foreign key from the monolithic tables, and for microservice approach we need a database per service, so the next section illustrate how can we decompose the monolithic database.

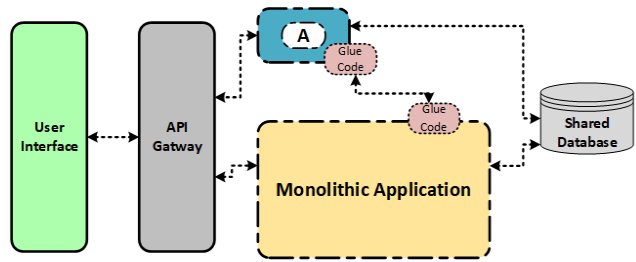

Figure 5: Shared database for monolithic and the extracted service

## 4. DECOMPOSE THE DATABASE

After extract the required service from the monolithic code base we need to compose the related database for that service, and to do so we need to analyse the database mappings and find out the related tables of the extracted service, this task – extracting the related tables from the database- is a complex task since because the separation between database objects may be not clear especially on relational database [13] because our target is no remove table joins between the extracted service database and the database of the monolithic application, and also there should be no hard constraints like foreign keys or database triggers between the two databases, and to partially achieve these targets we need to move all tables joints and databases constraints from the database layer to the business logic layer, so in case of the two databases has a shared identifier that uniquely identify database entries, this identifier should belongs to one database that is related to the service that manage the life cycle of the entity where the identifier exists, and all other services should use that identifier as a reference via API communications, where the main point in database decomposition is to prevent the new extracted service from access the monolithic application database directly and vice versa.

After identifying the related tables for the extracted service, we need to create a new table for it by mirroring the related tables structure from the monolithic database. This will keep the code simple and make the data migration simple.

The next step is to synchronize the existing tables with the new one as in figure 6, synchronization here is just like creating a read-only replica of extracted service related tables from the old database to the extracted service database, and after synchronization we point the new extracted service to the new table and if any other service need data from the new table it cannot access the new database directly, instead other services can access the new database through calling the extracted service via API and the extracted service process the request and reply with result via API.

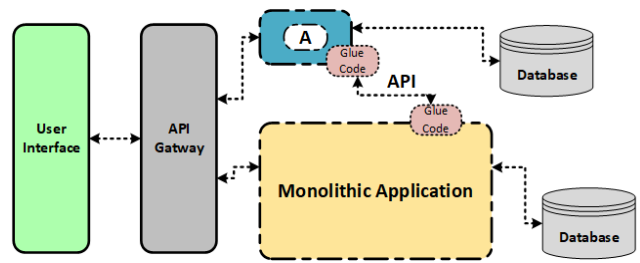

Figure 6: Decompose database

Create a new table then synchronizing the required tables from the old database leads to additional network and database calls.

## 5. CASE STUDIES

Many enterprises have adopted microservice architecture due to the increasing complexity of the modern application that can solved using the advantages of microservices architecture, as an example, Netflix is one of the early enterprise that convert their existing monolithic application to microservice design [14], Netflix launched in 1998 as a DVDs rent company, and in 2007 they introduced their video streaming service through monolithic web application which was hosted on their own datacenter, on August 2008 the monolithic application went down for three days because of a missing single semicolon that cause a corruption in their database.

In 2009, Netflix started to move from their own datacenter to Amazon Web Service "AWS", they firstly extracted movie encoding service which is a non-customer facing application to AWS, and in 2010 they moved the remaining service such as movie selection and accounts login/signup to AWS. They also replicate all the data from their single database to the new microservice database, and during the migration of customer-facing applications they faced latency issues on the web pages, and that was solved by managing resources within AWS and tuning AWS networking which has more variable latency than Netflix data centers.

Netflix consists of three components which are client interface, back-end and content delivery network "CDN" [15], where client interface is web browser or Netflix mobile application. Netflix back-end is all Netflix services and databases which run on AWS and back-end is responsible for most Netflix functionality except streaming videos, and the third part of Netflix is the open connect CDN that's distributed servers called Open Connect Appliance "OCA" that's located on some ISPs around the world where Netflix made partnership with these ISPs, these OCAs responsible for storing and streaming videos, Netflix distribute their OCAs as close as possible to the clients so that will reduce latency and improve users experience.

Netflix back-end shown in figure 7 that contains the microservice application, when client send request to Netflix application, AWS Elastic Load Balancer "ELB" that forward the request to API gateway service named Zuul [16] that developed by Netflix and has main role of traffic routing, traffic monitoring and security, Zull process traffic using pre-defined filters then forward it to application API which acts as orchestration layer of Netflix microservice, and it handle requests by call microservices in the desired order based on the request.

Since microservice can call each other's, Netflix developed Resilience4j that used for latency control, fault tolerance and controlling cascading failures by isolating each microservice from the caller process, and the result of each microservice can be cashed in memory so that in case any microservice takes long time to response Resilience can access the cash and return the last cashed result, so the latency would be minimal.

Also, Netflix back-end has stream processing pipeline for users' recommendation tasks and real time business intelligence tasks where results can be stored on Hadoop or AWS Simple Storage Service "S3".

Netflix developed Eureka that's has a main role of service discovery and registry, so when services are registered on Eureka, they can find each other dynamically, and in case of multiple instances of the same service are registered on Eureka, then Eureka route traffic to these services using round-robin manner.

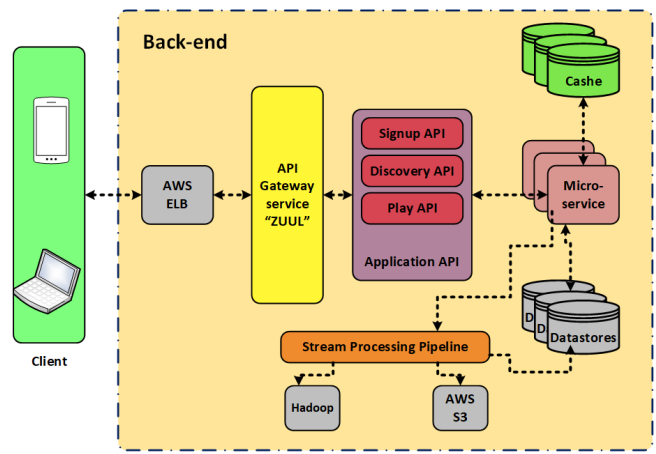

*Figure 7: Netflix Back-end Architecture*

When Netflix migrate to microservice and deployed it in AWS they used different types for databases as shown in figure 8, the use MySQL database for movie title management and ordering processes, while Casandra [17] database which is NoSQL database for process that has large number of read requests like logging, CDN routing and for real time data analysis they use Hadoop [18].

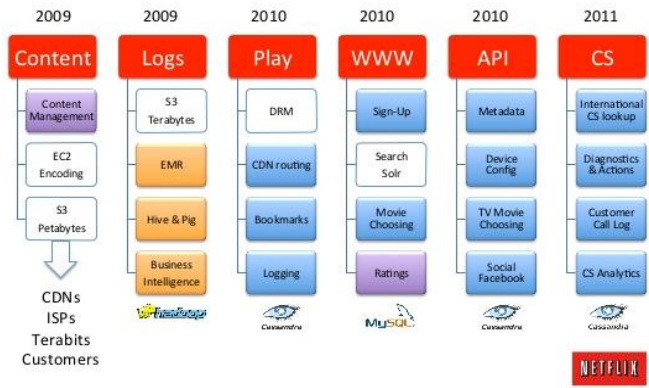

*Figure 8: Netflix databases deployed on AWS*

Netflix uses serverless for occasional services like video encoding that's When a video is uploaded to Amazon S3 it triggers a function for encoding the video into 60 different parallel streams, also they use serverless for file backup that's triggered when a file is changed or modified.

Netflix has developed many tools for developing and managing microservice applications, all these tools are put together in the Netflix Open-Source Software Centre.

Airbnb is another enterprise that converted their monolithic application to microservice their application was built using Ruby on Rails, and since they were growing rapidly, and they faced many incidents in their monolithic so their engineers took more time for incidents troubleshooting that developing new features, all that drove Airbnb [19] to use microservice architecture, they start the migration by building new services outside the monolithic application using strangler fig pattern, the first service they built outside the monolithic was the search engine so the application could more quickly respond to user queries based on date and destination, after that they build pricing service that is used for price prediction using machine learning models, also they built fraud protection service, after that they used SmartStack [20] that's an open-source automated service discovery framework which provided service registration, deregistration, health checking of services, and load balancing client traffic. After that, Airbnb engineers extracted "Product logics" out of the monolithic and finally they extracted all other services out of the monolithic.

## 6. CONCLUSION

In this paper we introduced how to convert the existing monolithic application into microservice application by firstly convert the base code using domain driven design which allow us to take the right decision to choose which service to be extracted first, after extracting the desired service we show how to split the monolithic database and extracted the related tables for the extracted service to be as stand-alone database that can be accessed privately using the service only.

During the migration developers will face a lot of challenges and issues in terms of extracted service context which can be overcome using domain-driven design, the relational database extraction that can be solved by moving all tables joints and databases constraints from the database layer to the business logic layer, and another issue that developers need to handle is that since microservice deployed on cloud services like Amazon, Google or Azur, that implies developer to be familiar with cloud computing.

Developers can also take advantage of how large enterprises such as Netflix or Airbnb converted their existing monolithic application into an agile microservice application.

Also we can combine microservice architecture with serverless to get serverless microservice, which is a microservice design with some services that run occasionally and are deployed as an event-driven function.